\documentclass[aps,prc,reprint,groupedaddress]{revtex4-1}
\usepackage{epsfig}
\usepackage{multirow}
\begin{document}


\title{Reply to comments on `Structure effects in the $^{15}$N($n,\gamma$)$^{16}$N radiative capture reaction from the Coulomb dissociation of $^{16}$N'}


\author{Shubhchintak}
\email{shub.shubhchintak@tamuc.edu}
\affiliation{\it {Department of Physics and Astronomy, Texas A$\&$M University - Commerce, 75429, USA}}
\author{Neelam}
\email{nph10dph@iitr.ac.in}
\affiliation{Department of Physics, Indian Institute of Technology - Roorkee, 247667,
INDIA}
\author{R. Chatterjee}
\email{rcfphfph@iitr.ac.in}
\affiliation{Department of Physics, Indian Institute of Technology - Roorkee, 247667,
INDIA}


\date{\today}


\maketitle

The purpose of our paper \cite{ref1} was to primarily bring into focus a long standing controversy on the spectroscopic factors of $^{16}$N states, which would have consequences on the $^{15}$N($n,\gamma$)$^{16}$N capture reaction. Our calculations seem to favor the spectroscopic factors from Ref. \cite{ref2} against those of Refs. \cite{ref3,ref4}.

Our Coulomb dissociation theory is formulated within the post-form finite range distorted wave Born approximation where the electromagnetic interaction between the fragments and the target nucleus is included to all orders and the breakup contributions from the entire non-resonant continuum corresponding to all the multipoles and the relative orbital angular momenta between the fragments are accounted for. Given that the only input to our theory is the ground state wave function of the projectile, of any orbital angular momentum configuration, makes our method  free from the uncertainties associated with the multipole strength distributions occurring in many other formalisms which require the exact positions and widths of the continuum states. This is precisely the reason why the potential model referred to in Ref. \cite{ref5} would have problems in deciding an unambiguous criterion to construct the continuum states. While this is well known in the literature, we refer only to one reference \cite{ref6} especially in the calculation of transition strengths in (core + nucleon) nuclei.

For the sake of completeness, we have also plotted the branching ratio for the $^{15}$N($n,\gamma$)$^{16}$N capture reaction in our case (defined as the ratio of cross section to a given state divided by the total cross section) against the center of mass energy (${\rm{E_{c.m.}}}$), in Fig. 1.
Indeed the branching ratios are energy dependent. 
\begin{figure}[t] 
\centering
\includegraphics[height=6.0cm, clip,width=0.35\textwidth]{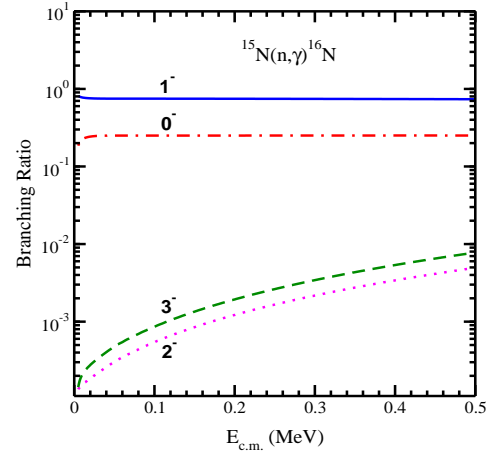}
\caption{\label{fig:1}(Color online) The branching ratio to various states of $^{16}$N. Dotted and dashed lines  correspond to $2^-$ and $3^-$ states, respectively. Dot-dashed and solid lines correspond to $0^-$ and $1^-$ states, respectively.}
\end{figure}

In fact, it is now known that if one treats a many-body nuclear system as an open quantum system, wherein the bound states and the  many-body continuum of decaying states are treated together on the same footing, spectroscopic factors or more precisely single-nucleon overlap integrals can become energy dependent \cite{refa,refb,refc}. Since the post-form reaction theory in our case includes the entire non-resonant continuum, we believe, could be the reason for energy dependent branching ratios. It is also interesting to note that this energy dependence is sensitive to the relative orbital angular momentum content of the state. The $2^-$ and $3^-$ states, which have energy dependent branching ratios, essentially contain $d-$wave neutrons while the $0^-$ and $1^-$ states, which have energy independent branching ratios and also accounts for almost all of the total cross section, contain $s-$wave neutrons (see also Table I of Ref. \cite{ref1}). However, this issue needs to be investigated in more details.

The next question one has to confront is that if the theory \cite{ref1} includes the “entire non-resonant” contribution then how can one justify the use of a relation like Eq. (3) in Ref. \cite{ref1} to extract the photo-dissociation cross section from the relative energy spectra. This is only possible if and only if a single multipole dominates, as we have checked from Ref. \cite{ref7} (quoted as Ref. [13] in \cite{ref1}). 
	
Finally, given the paucity of data for the $^{15}$N($n,\gamma$)$^{16}$N at low energies indirect approaches, like the Coulomb dissociation, may be helpful although we certainly agree that they would be difficult.  We again reiterate that we have attempted to resolve the discrepancy in the spectroscopic factors of low-lying $^{16}$N levels and that it is essential to know the low energy $^{15}$N($n,\gamma$)$^{16}$N capture cross section, especially below 0.25 MeV.

\section*{Appendix}

In Fig. 2 of Ref. \cite{ref1}, the data points in the top panel appear to be a bit shifted as they were inadvertently plotted against the laboratory energy instead of the center of mass energy. We now reproduce the correct Fig. 2 of Ref. \cite{ref1}. This change does not alter any of the discussions or conclusions of the work.

\begin{figure}[ht]
\centering
\includegraphics[height=8.5cm, clip,width=0.45\textwidth]{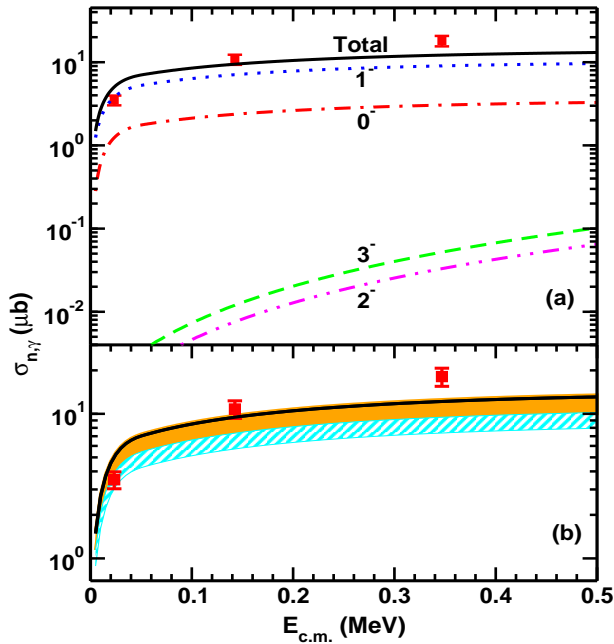}
\caption{\label{fig:2} (Color online) (a) Total non-resonant $^{15}$N$(n,\gamma)^{16}$N cross section (solid line) obtained by summing up contributions of capture to all four states of $^{16}$N (given in Table I of Ref. \cite{ref1}) using their respective shell model $C^2S$. (b) Total non-resonant capture cross section obtained by using the experimentally extracted $C^2S$ (including uncertainties) from Ref. \cite{ref2} (filled band) and Ref. \cite{ref3} (filled pattern) compared with the total non-resonant cross section (solid line) shown in (a). The experimental data in both panels are from \cite{meissner}.}
\end{figure}

\section*{Acknowledgments}
This text results from research supported by the Department of Science and Technology, Govt. of India, (SR/S2/HEP-040/2012). [S] is supported by the U.S. NSF Grant No. 1415656 and the U.S. DOE Grant No. DE-FG02-08ER41533.


\end{document}